\documentclass[journal=nalefd,manuscript=article]{achemso}

\usepackage{chemformula} 
\usepackage[T1]{fontenc} 
\usepackage{booktabs}
\usepackage{hyperref}
\usepackage{orcidlink}
\usepackage{svg}
\usepackage[dvipsnames]{xcolor}

\usepackage{xspace}
\newcommand*{\MS}{MoS$_2$\xspace}
\newcommand*{\WS}{WS$_2$\xspace}
\newcommand*{\EF}{E$_F$\xspace}

\author{Surender Kumar\orcidlink{0009-0000-3072-5633}}
\affiliation{Institut für Festk\"orpertheorie und -Optik, Friedrich-Schiller-Universit\"at Jena, 07743 Jena, Germany}
\email{surendermohinder@gmail.com}
\altaffiliation{These authors contributed equally to this work.}

\author{Markus Fröhlich\orcidlink{0000-0001-6543-0843}}
\affiliation{Institute of Physical and
Theoretical Chemistry, Eberhard Karls University of Tübingen, 72076 Tübingen, Germany}
\altaffiliation{These authors contributed equally to this work.}

\author{Stefan Velja\orcidlink{0009-0003-1268-6273}}
\affiliation{Institut für Festk\"orpertheorie und -Optik, Friedrich-Schiller-Universit\"at Jena, 07743 Jena, Germany}

\author{Marco Kögel\orcidlink{0009-0004-5065-1378}
}
\affiliation{NMI Natural and Medical Sciences Institute at the University of Tübingen, 72770 Reutlingen, Germany}
\alsoaffiliation{University of Tübingen, Institute of Applied Physics, 72076 Tübingen, Germany}

\author{Onno Strolka\orcidlink{0000-0002-1096-5610}}
\affiliation{Institute of Physical and
Theoretical Chemistry, Eberhard Karls University of Tübingen, 72076 Tübingen, Germany}
\alsoaffiliation{Leibniz University of Hannover, Cluster of Excellence PhoenixD (Photonics, Optics and Engineering - Innovation Across Dimensions), 30167 Hannover, Germany}

\author{André Niebur\orcidlink{0000-0001-8834-303X}}

\affiliation{Leibniz University of Hannover, Cluster of Excellence PhoenixD (Photonics, Optics and Engineering - Innovation Across Dimensions), 30167 Hannover, Germany}

\author{Samuell Ginzburg\orcidlink{0009-0002-2786-3503}}
\affiliation{Yusuf Hamied Department of Chemistry, University of Cambridge, Lensfield Road, Cambridge, UK}

\author{Muhammad Sufyan Ramzan\orcidlink{0000-0002-5017-8718}}
\affiliation{Institut für Festk\"orpertheorie und -Optik, Friedrich-Schiller-Universit\"at Jena, 07743 Jena, Germany}

\author{Jannik C. Meyer\orcidlink{0000-0003-4023-0778}}
\affiliation{University of Tübingen, Institute of Applied Physics, 72076 Tübingen, Germany}
\alsoaffiliation{NMI Natural and Medical Sciences Institute at the University of Tübingen, 72770 Reutlingen, Germany}

\author{Jannika Lauth\orcidlink{0000-0002-6054-9615}}

\affiliation{Institute of Physical and
Theoretical Chemistry, Eberhard Karls University of Tübingen, 72076 Tübingen, Germany}
\alsoaffiliation{Leibniz University of Hannover, Cluster of Excellence PhoenixD (Photonics, Optics and Engineering - Innovation Across Dimensions), 30167 Hannover, Germany}

\email{jannika.lauth@uni-tuebingen.de}

\author{Caterina Cocchi\orcidlink{0000-0002-9243-9461}}
\affiliation{Institut für Festk\"orpertheorie und -Optik, Friedrich-Schiller-Universit\"at Jena, 07743 Jena, Germany}
\email{caterina.cocchi@uni-jena.de}

\title{Atomistic Origin of Photoluminescence Quenching in Colloidal \MS and \WS Nanoplatelets}


\begin{document}

\begin{tocentry}
\centering
\includegraphics[]{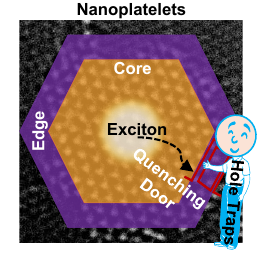}
\end{tocentry}

\newpage
\begin{abstract}
Large chemical tunability and strong light-matter interactions make colloidal transition metal dichalcogenide (TMD) nanostructures particularly suitable for light-emitting applications. However, ultrafast exciton decay and quenched photoluminescence (PL) limit their potential. Combining femtosecond transient absorption spectroscopy with first-principles calculations on \MS and \WS nanoplatelets, we reveal that the observed sub-picosecond exciton decay originates from edge-located optically bright hole traps. These intrinsic trap states stem from the metal $d$-orbitals and persist even when the sulfur-terminated edges are hydrogen-passivated. Notably, \WS nanostructures show more localized and optically active edge states than their \MS counterparts, and zigzag edges exhibit a higher trap density than armchair edges. The nanoplatelet size dictates the competition between ultrafast edge-trapping and slower core-exciton recombination, and the states responsible for exciton quenching enhance catalytic activity. Our work represents an important step forward in understanding exciton quenching in TMD nanoplatelets and stimulates additional research to refine physicochemical protocols for enhanced PL. 
\end{abstract}

\newpage
Substrate-free colloidal synthesis is a scalable and versatile bottom-up method to produce transition metal dichalcogenide (TMD) nanostructures.~\cite{Mahler2014,Sun2017,Sun2021,Chen2015,Zechel2023, Kapuria2023}. This approach offers thermodynamically stable 2H-phase monolayers~\cite{Zechel2023} with lateral sizes down to a few nanometers~\cite{Niebur2023,Fröhlich2024,JOSEPH2023,Chen2015,Zechel2023}, enabling fine control over composition and surface chemistry~\cite{Mahler2014,Sun2017,Sun2021,Fröhlich2024,JOSEPH2023,Chen2015,Zechel2023}. Among low-dimensional TMDs, \MS and \WS stand out for their direct visible-range bandgaps, strong light-matter coupling, and tunable optical properties~\cite{Gopalakrishnan2015,Wang2016,NGUYEN2019,GOLOVYNSKYI2021,Doolen1998,Gan2015,Li2017,Mani2020,Yin2018,Chiu2023,Zechel2023,BERTRAM2006,mak2010,Wang2016,wang2012}, which makes them highly promising for thin-film integration in optoelectronic devices.

Bright photoluminescence (PL), long-lived emissive states, and tunable emission spectra are highly desirable properties for light-emitting applications. TMDs exhibit these optimal optical features thanks to their exceptionally bound excitons ($\sim$0.5~eV binding energies)~\cite{Chernikov2014,mak2010}. 
Although excitons are expected to radiatively recombine with high quantum efficiency~\cite{Brokmann2004,Nirmal1996,Durisic2009,Banin1999}, colloidal TMD nanostructures typically show weak or quenched PL~\cite{Frauendorf2024,Pippia2022-2,Pippia2023,Schiettecatte2019, Markus2025} compared to monolayers obtained via exfoliation/chemical vapor deposition (CVD).~\cite{Cunningham2016,li2014,Eda2011,Valeria2013,Lee2012,zhao2013}
This PL loss can originate from non-radiative recombination at surface or edge sites~\cite{Gopalakrishnan2015,Wang2016,NGUYEN2019,GOLOVYNSKYI2021}, charge carrier trapping~\cite{Schiettecatte2019,Frauendorf2024,Doolen1998,Pippia2023, Mukherjee2023,Wang2015,Cunningham2016}, and insufficient chemical passivation of surface states~\cite{Zhang2018,Gan2015,Li2017}. However, the intrinsically disordered nature of colloidal samples makes it non-trivial to isolate the specific physical contributions. As a result, the fundamental, atomistic origin of PL quenching and charge carrier trapping in colloidal TMD nanostructures remains unclear to date.

\textit{Ab initio} structure calculations represent a valuable resource to complement structural analysis and spectroscopic characterization of colloidal systems. Density functional theory (DFT) has been successfully applied to determine trap states in III-V and II-VI colloidal nanostructures~\cite{Houtepen2017,Giansante2017,Kirkwood2018,Alexander2024,Kumar2025}, revealing their microscopic origin and electronic character. These exemplary results demonstrate the potential of this parameter-free approach, which can contribute to accurately identify the microscopic origin of PL quenching in colloidal TMD nanostructures, by systematically investigating different compositions, sizes, and terminations with atomistic control. 

By combining transient absorption spectroscopy with DFT calculations, we identify edge-localized metal $d$-orbitals acting as trap states to be responsible for ultrafast exciton decay in colloidal \WS and \MS nanoplatelets (NPLs), driving sub-picosecond PL quenching. Furthermore, we reveal how the optical properties of TMD NPLs evolve with size and how their edge states, while detrimental to PL, are key to their catalytic activity. Our findings establish a clear link between material morphology, fundamental photophysics, and functional performance, providing a rational basis for the design of versatile wet-chemically synthesized TMD nanostructures.

\begin{figure}[!ht]
    \centering
\includegraphics[width=0.5\textwidth]{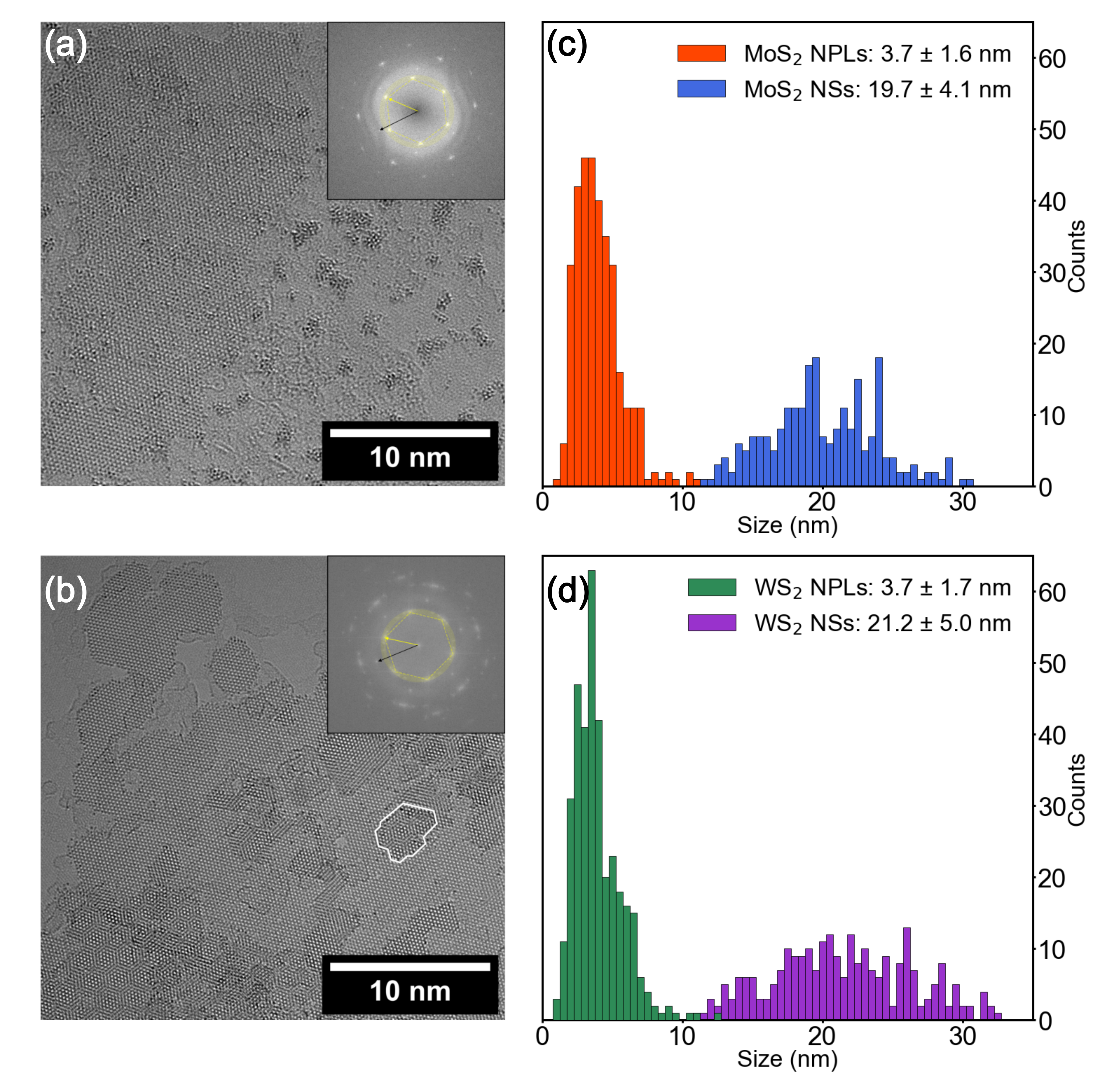}
    \caption{High resolution TEM-image of (a) \MS NSs and (b) \WS NPLs, different in shape, size and orientation. A single exemplary NPL is highlighted with a white border. The hexagonal crystal structure of the semiconducting 2H-phase can be seen in the FFT-inset (yellow) with the underlying graphene structure indicated by the black arrow. Size distributions of (c) \MS, and (d) \WS NPLs and NSs respectively.} 
    \label{fig1}
\end{figure}

\MS and \WS colloidal NPLs and nanosheets (NSs) [Figure~\ref{fig1}] are synthesized by dropwise addition of MoCl$_5$/WCl$_6$ dissolved in oleylamine (OlAm) into a 320~$^{\circ}$C hot solution of elemental sulfur and hexamethyldisilazane (HMDS) in OlAm.\cite{Mahler2014} A short addition time of 10~min and using sulfur in excess ($\sim$700~equivalents (eq)) limits the lateral growth, resulting in NPLs.~\cite{Niebur2023}. A longer addition time and a lower sulfur excess (3.4~eq) on the other hand yields laterally larger NSs.~\cite{Niebur2023, Kapuria2023, Pippia2022, Fröhlich2024} Due to agglomeration caused by solvent evaporation, the NPLs arrange in macroscopic structures, resulting in the observed Moir\'e patterns. OlAm ligands used in the synthesis ensure the spatial separation of the NPLs/NSs even when agglomerated, and has been reported previously.~\cite{Frauendorf2021,Fröhlich2024,Markus2025} \MS NPLs have an average lateral size of 3.7~$\pm$~1.6~nm and NS a size of 19.7~$\pm$~4.1~nm~[Figure~\ref{fig1}(c)]. \WS NPLs and NSs [Figure~\ref{fig1}(d)] show similar sizes (3.7~$\pm$~1.7~nm and 21.2~$\pm$~5.0~nm). Both NPLs and NSs have a hexagonal 2H-phase crystal structure, highlighted in the fast Fourier transform (FFT)-inset of Figure~\ref{fig1}(a,b). 

{The semiconducting 2H-phase purity and monolayer nature of these TMDs have been confirmed in previous studies by means of X-ray photoelectron spectroscopy, X-ray diffraction, and Raman spectroscopy, showing no evidence of metallic polytypes.\cite{Niebur2023, Fröhlich2024} Our established annealing protocol ensures a quantitative transition to the 2H-phase even in high-surface-area NPLs.\cite{Niebur2023} The monolayer character of the samples is further corroborated by excitonic peak positions and the observation of second-harmonic generation.\cite{Markus2025} While individual AFM thickness measurements are complicated by the propensity of these NPLs to form ligand-separated ``nanoflower'' aggregates upon drying, the ensemble structural and optical fingerprints are consistent with high-quality monolayers.\cite{Markus2025}}

Compared to the larger NSs, the reduced size of NPLs introduces significant lateral confinement,\cite{Frauendorf2024} resulting in a hypsochromic shift of the excitonic resonances. In \MS the A/B excitons shift from 1.91/2.06~eV (649/601~nm) in NSs to 2.04/2.18~eV (609/568~nm) in NPLs. In contrast, the shifts in \WS are are 2.02/2.39~eV (614/518~nm) in Nss and 2.04/2.45~eV (607/506~nm) in NPLs (see Figure~S1 and Table~S1). Notably, the A exciton of \MS exhibits a much larger shift (130~meV, 40~nm) compared to \WS (20~meV, 7~nm), while also showing a more pronounced decrease in intensity. This indicates that the same lateral confinement to NPLs more prominently affects the electronic structure of \MS. Similarly, in earlier work, we discussed the vanishing absorption of the A excitonic transition in colloidal \MS NPLs, showing that the dominant feature at 2.09~eV (593~nm) originates from the B-exciton, shifted from  2.03~eV (610~nm) due to lateral confinement.~\cite{Niebur2023,Frauendorf2024} The broad absorption is attributed to a combination of inhomogeneous size broadening and morphological variations [Figure~\ref{fig1}(a,b)], since both the relative size distribution and the edge-to-core atom ratio are higher in NPLs than in the NSs. The reduced absorption intensity in NPLs is a result of the smaller NPLs is a direct consequence of reduced oscillator strength, which scales with the lateral surface dimensions.~\cite{Fouladi-Oskouei2018,Ayari2020,Roda2022} 

 \begin{figure}
    \centering
\includegraphics[width=0.95\textwidth]{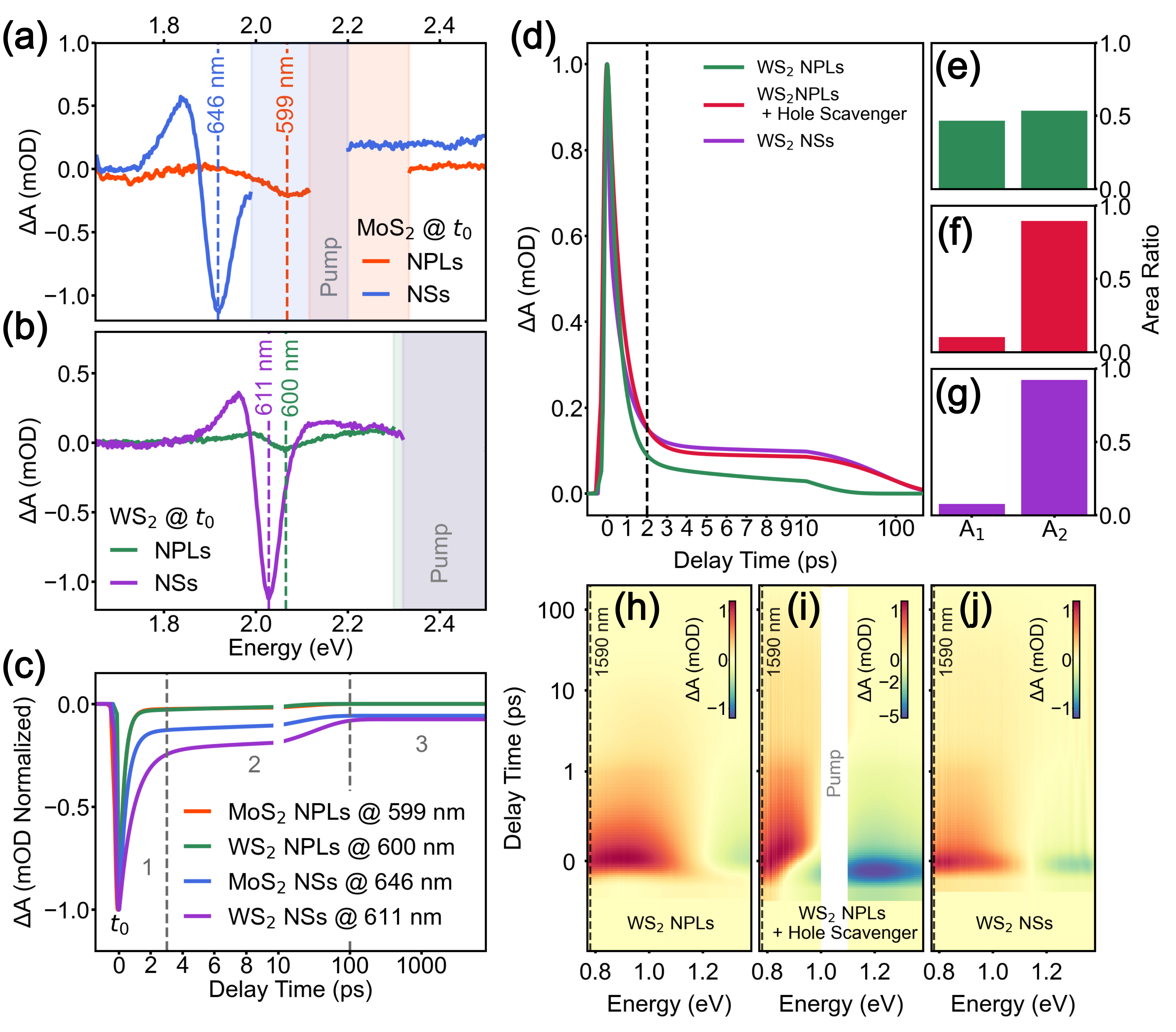}
    \caption{(a, b) Spectral line cuts of \MS (a) and \WS (b) at $t_0$ showcasing shift, broadening and weakening of the A excitonic transition under lateral confinement. The spectral position of each A exciton is marked by a vertical line, while the areas obscured by the pump pulse are marked in the resp. color of the corresponding graph. (c) Fitted decay dynamics of the A exciton of all samples. The time scale is linear between -0.5 and 10~ps and logarithmic from 10-7500~ps to allow both high resolution of the early response and maximum coverage. {(d) Fitted decay traces of \WS NPLs (green), \WS NPLs with ascorbic acid added as hole scavenger (red) and \WS NSs (purple), Integrated areas under each trace assigned to the first and second decay process (A$_1$, A$_2$) for (e) \WS NPLs, (f) \WS NPLs and hole scavenger and (g) \WS NSs, showing a reduced contribution of the initial trapping (first component) when the hole scavenger is present. Associated hyperspectra showing \WS NPLs (h), \WS NPLs with hole scavenger (i) and \WS NSs (j).}}
    \label{fig2}
\end{figure}

{The transient absorption (TA) spectra, acquired immediately after photoexcitation (t$_0$) [Figure~2(a,b)], reproduce the trends observed in steady-state measurements, including the hypsochromic shift for \MS and \WS NSs/NPLs of 130~meV/40~nm and 20~meV/7~nm, respectively, and spectral broadening. To probe the dynamics, all samples were excited at their respective B-exciton energy (details in Figure~S1) under a low excitation density of 4.7~$\mu$J/cm$^2$. The resulting carrier dynamics were extracted using global analysis of a sequential bi/tri-exponential model. This approach allows us to independently resolve decay times from overlapping signals while accounting for the instrument response function, ensuring accurate characterization of the sub-picosecond dynamics near the resolution limit.~\cite{Frech2025} (further described in the SI). As shown in Figure~\ref{fig2}(c), the A exciton bleaches of \MS and \WS NPLs exhibit a significantly faster decay compared to the larger NSs (see Table~\ref{tab1}).}

\begin{table}
\caption{{Decay times fitted by a sequential 
bi/tri-exponential model for initial carrier trapping $t_1$, band gap renormalization $t_2$, and slow recombination from trapped states $t_3$ of \MS and \WS NPLs/NSs.}}
\begin{tabular}{c|c|c|c} 
 & t$_1$ & t$_2$  & t$_3$    \\ \hline
\MS NPLs & (270 $\pm$ 50) fs  & (21 $\pm$ 7) ps & -  \\ 
\MS NSs &  (410 $\pm$ 100) fs  & (40 $\pm$ 7) ps & $\rightarrow$ $ \infty$ \\
\hline
\WS NPLs & (291 $\pm$ 40) fs  &(11 $\pm$ 1) ps & -  \\ 
\WS NSs &  (741 $\pm$ 40) fs  & (29 $\pm$ 2) ps & $\rightarrow$ $\infty$
\label{tab1}
\end{tabular}
\end{table}

A rapid, sub-picosecond trapping process is observed in both \MS and \WS nanostructures. However, as shown in Table~\ref{tab1}, this effect is more pronounced in the NPLs, being faster and contributing more to the overall decay. Such a behavior points to a higher density of mid-gap trap states favoring non-radiative decay in NPLs.~\cite{Niebur2023, Frauendorf2024}. In monolayer colloidal TMD NSs, electron trapping mechanisms have been linked to sulfur vacancies~\cite{Pippia2023}, whereas hole trapping has been reported for multilayer NSs~\cite{Schiettecatte2019}. These similar decay mechanisms cannot be directly translated to interpret the results obtained for smaller \MS and \WS NPLs. 

The reduced lateral dimensions of NPLs lead to a significantly higher density of exposed edge sites compared to NSs, fundamentally altering exciton decay pathways. {While both nanostructures show picosecond bathochromic absorption shifts consistent with band gap renormalization~\cite{Schiettecatte2019,Pippia2023,Pogna2016,Frauendorf2024} (Figure~S2), a distinct, long-lived process assigned to radiative recombination, is observed exclusively in the larger NSs [Figure~\ref{fig2}(c)], where $\mu$-PL has been previously reported~\cite{Frauendorf2021, Pippia2022-2, Markus2025}. In contrast, NPL dynamics are dominated by sub-picosecond trapping, which our measurements suggest is mediated by intrinsic mid-gap states. This accelerated non-radiative pathway effectively outcompetes radiative recombination, rationalizing the absence of detectable photoluminescence in the NPLs compared to their larger NS counterparts.}

{To isolate these carrier dynamics from ground-state absorption artifacts (Figure~S3), we performed selective photoexcitation of the A and B transitions in \WS and probed the near-infrared region. We observe a broad, positive excited-state absorption (ESA) signal [Figure~2(h-j)] assigned to intraband transitions~\cite{Wang2015TA}, where the initial decay component reflects the trapping process observed in the visible range~\cite{Schiettecatte2019}. Kinetic analysis along the 0.78~eV probe line reveals that trapping in NPLs is consistently faster ($\Delta t_1 \approx 20$~fs) upon A-exciton excitation, suggesting accelerated hole trapping when carriers are generated near the band edge (Figure~S4). To confirm the hole-mediated nature of this process, we utilized ascorbic acid as a hole scavenger~\cite{Dunklin2018}. The presence of the scavenger significantly extends the ESA decay lifetime [Figure~2(d)] and reduces the trapping contribution from 48\% to $\sim$10\% in NPLs [Figure~2(e-g)]. These findings conclusively identify holes as the primary drivers of the sub-picosecond trapping dynamics in \WS NPLs.}

\begin{figure}[!ht]
    \centering
\includegraphics[width=0.925\textwidth]{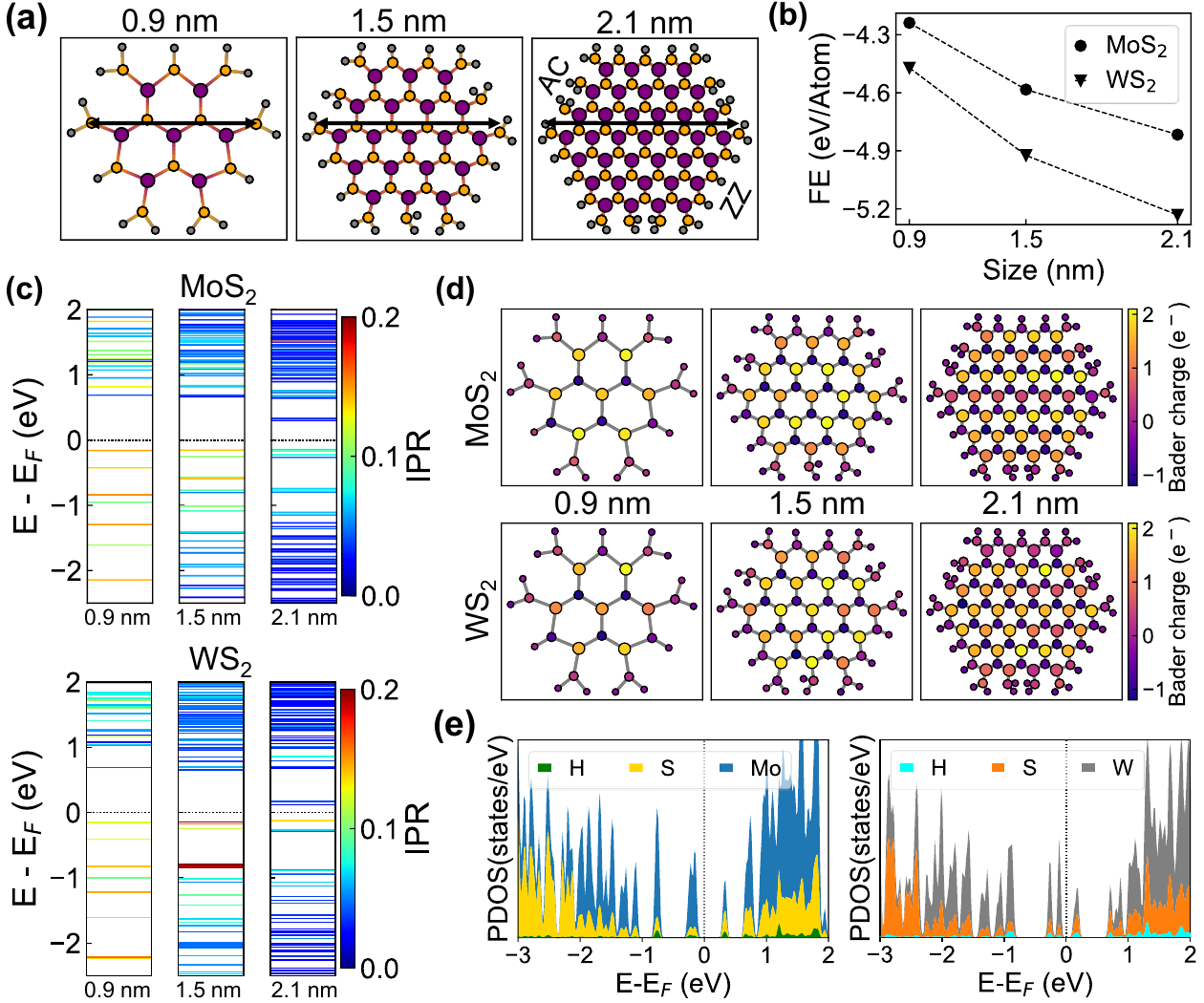}
\caption{(a) Optimized structures of hydrogen-passivated \MS\ nanoplatelets (NPLs) of different sizes with zigzag (ZZ)and armchair (AC) S-edges. Mo atoms are shown in purple, S in yellow, and H in gray. Black arrows indicate the size. (b) Formation energies of \MS\ and \WS\ NPLs as a function of size.
(c) Inverse participation ratio for states around Fermi energy (\EF ) for \MS and \WS NPLs.
(d) Net Bader charges for calculated \MS (top) and \WS (bottom) NPL sizes.
(e) Projected density of states (PDOS) near the Fermi level for 2.1 nm \MS (left) and \WS\ (right) NPLs.}
\label{fig3}
\end{figure}

To understand the mid-gap states inferred from our time-resolved spectroscopic measurements, we perform DFT calculations to investigate the electronic structure of a series of \MS and \WS NPLs. Our model structures [Figure~\ref{fig3}~(a)] are monolayer-thick pseudo-hexagonal 2H-polymorphic NPLs with varying lateral dimensions of 0.9, 1.5, and 2.1~nm. These model structures are qualitatively representative of the experimental NPLs reported in Figure~\ref{fig1}~(a). Their size falls in the experimental range, which is reasonable for rationalizing the measurements. The optimized geometries are defect-free and fully passivated, featuring zigzag (ZZ, edge S-atoms bonded to two Mo (W) atoms) and armchair (AC, edge S-atoms bonded to a single Mo (W) atom) edges. {Hydrogen passivation effectively removes spurious mid-gap states from under-coordinated S-atoms, while preserving the dominant electronic characteristics dictated by the edge species~\cite{krumland2021exploring}. It is also worth noting that H-terminated colloidal TMD nanostructures have been reported in the experimental literature~\cite{Seyyedmajid2018,Lauritsen2003,Topsoe1993,Kyung2017,Din2021}. }

 The formation energy [Figure~\ref{fig3}~(b)] decreases as both \MS and \WS NPL size grows. This trend is consistent with the size distribution observed in our measurements [Figure~\ref{fig1}~(c)] and reflects the enhanced thermodynamic stability of larger NPLs. At comparable sizes, \WS NPLs are consistently more stable than \MS, due to the stronger bonds and higher ionic character of W-S relative to Mo-S bond~\cite{PANDEY2019}. 
 
To verify whether trap states are present in the NPLs, we investigate the electronic structure [Figure~\ref{fig3}~(c)] near the Fermi level (\EF). Both \MS\ and \WS\ NPLs show a semiconducting behavior with a band gap decreasing as the NPL size increases, following the expected trend of quantum confinement. The single-particle-state analysis based on the inverse participation ratio~\cite{Calixto2015, Kumar2025,Steenbock2024}[IPR, Figure~\ref{fig3}~(c)], a standard metric to assess wavefunction localization~\cite{Niebur2021,Niebur2021b}, we find that several hole (occupied) states near \EF in both \WS and \MS have high IPR values across all considered sizes, indicating a strong localization. In the smallest structures (0.9~nm) of both materials, most hole states are relatively localized compared to the largest NPL (2.1~nm). Conversely, the deeper occupied states become increasingly delocalized, approaching a bulk-like character. This trend is a size effect, resulting from the decreased relative number of edge atoms in larger flakes. In contrast, the electron (unoccupied) states remain largely delocalized, with only occasional localized states in the largest \MS and \WS NPLs. Notably, \WS NPLs show more localized states near \EF than \MS, which correlates with the differences in the temporal measurements {of the energy shift. The strongly localized states in \WS have energies that are only weakly dependent on their lateral size, resulting in a much smaller shift of 20 meV.}

\begin{figure}[!ht]
    \centering
\includegraphics[width=.9\textwidth]{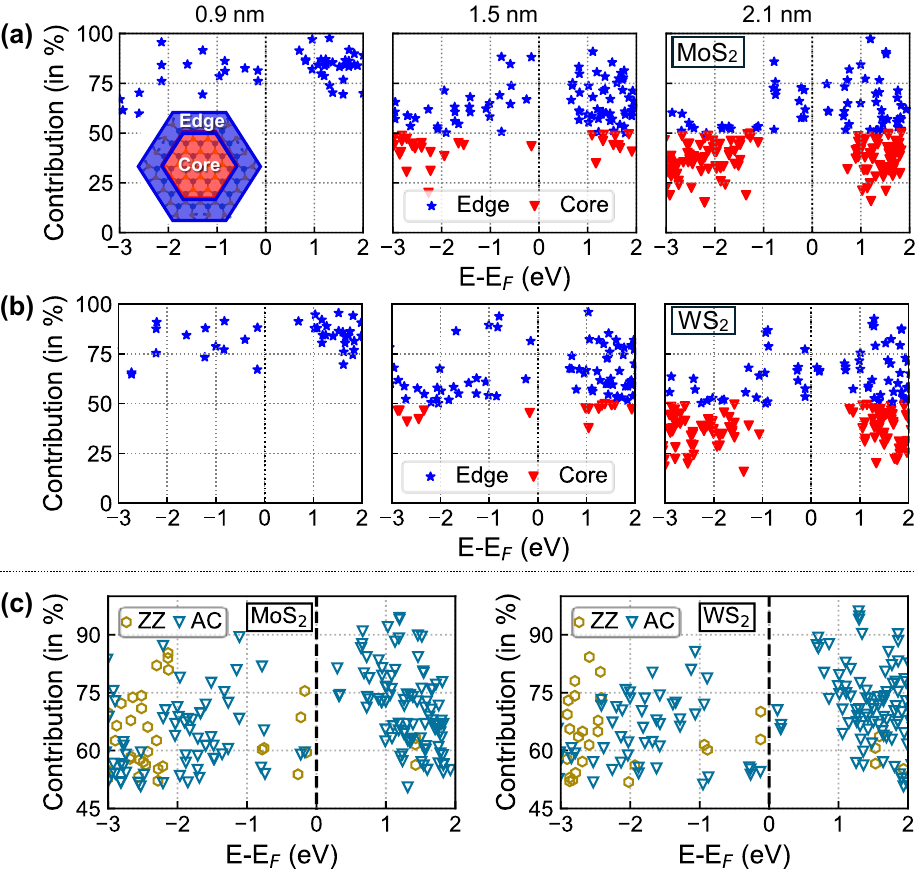}
\caption{Calculated core-edge contributions to energy levels near the Fermi energy in (a) \MS and (b) \WS NPLs. Blue stars (red triangles) mark states with $>50\%$ edge (core) contribution for the core-edge partitioning. (c) Relative contributions of zigzag (ZZ) and armchair (AC) edges to electronic states near the Fermi energy (\EF) in 2.1 nm \MS (left) and (b) \WS (right) NPLs.} 
    \label{fig4}
\end{figure} 

To analyze the origin of electronic state localization, we perform Bader charge analysis of local charge fluctuations in the considered structures [Figure~\ref{fig3}~(d)]. Each NPL has a uniform core and a chemically distinct edge (Mo-S-H or W-S-H), with charge deviations increasing in larger flakes [Figure~\ref{fig3}~(d)]. In the central region, electron transfer from Mo (or W) to S occurs, consistent with their different electronegativity. At the edges, disrupted bonding reduces the ability of atoms to fully share or accept electrons and breaks the local lattice periodicity, giving rise to spatially localized electronic states [Figure~\ref{fig3}~(c)] near \EF. Despite S-termination and H-passivation, edge metal atoms dominate the charge distribution, indicating their major role in the localized states.
Different charge fluctuation patterns emerge in \MS and \WS NPLs of equal size. The central metal atoms in \WS display a more uniform charge distribution, in contrast to the uneven charge distribution in \MS, which also leads to charge differences at the edges. 
The projected density of states (PDOS) of \MS\ and \WS\ NPLs provides additional information about the orbital contributions to the energy levels. As shown in Figure~\ref{fig3}~(e) for the systems with 2.1~nm size, the hole (occupied) states up to about 2~eV below the Fermi level are dominated by metal $d$-orbitals (Figure~S5), while S $p$-orbitals contribute deeper in energy, although signs of hybridization are present everywhere. In the electron (unoccupied) region, the trend does not change significantly, with the metal $d$-orbitals remaining predominant in the considered energy range. Here, hydrogen states contribute more than in the hole region.


We complete our analysis by inspecting the contributions from the interior (core) and exterior (edge) region of the considered NPLs [Figure~\ref{fig4}]. Specifically, we identify the origin of a specific state if more than 50\% of its contributions come from either portion of the nanostructure. Across all structures and in both compositions, both hole and electron states near \EF are predominantly edge-localized across all sizes of \MS and \WS [Figure~\ref{fig4}~(a,b)]. In the smallest flake (0.9~nm), where nearly all atoms lie at the periphery, all states obviously originate from the edges. As the flake size increases, core-localized states gradually emerge; however, even in the largest flake (2.1 nm), states within $\sim$1 eV of the Fermi level remain mostly edge-localized.

 For the largest 2.1~nm NPLs, we additionally investigate the contributions of states belonging to each edge type [Figure~\ref{fig4}~(c)]. Near \EF in the occupied region, states on zigzag edges contribute slightly more than those on armchair edges, while deeper hole states show increasing contributions from armchair edges. Electron states up to $\sim$1~eV are mostly dominated by armchair-edges contributions in both \MS and \WS NPLs. This edge-dependent behavior, characterized by localized states and charge imbalances, establishes the edges as chemically active sites. These findings thus explain why \MS and \WS nanostructures are highly effective in photocatalysis~\cite{Yuan2014,Kayal2024,Rahman2022,Li2018,Yuan2015,Jin2016,Xiao2018,Zou2018}.
 
\begin{figure}[!ht]
    \centering
\includegraphics[width=\textwidth]{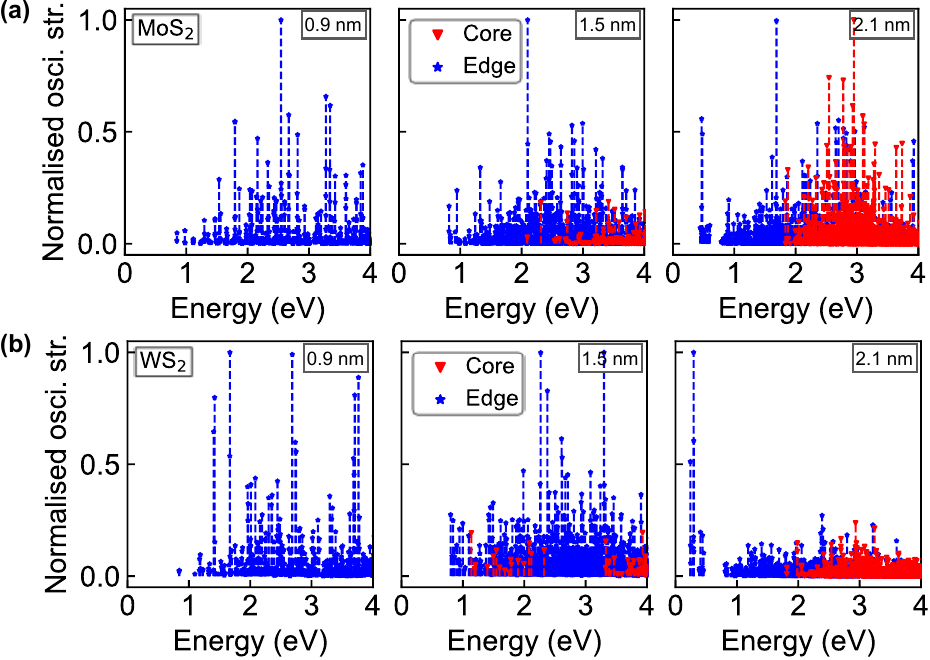}
   \caption{Calculated oscillator strengths for transitions below 4 eV in (a) \MS and (b) \WS NPLs, colored based on state character: red (core-originated, both states $<50\%$ edge character), blue (edge-originated, at least one state has $>50\%$ edge character).}
  \label{fig5}
\end{figure} 

To gain qualitative insight into the optical activity of the investigated NPLs, we calculate the normalized oscillator strengths within the independent-particle approximation (IPA), see Figure~\ref{fig5}. Consistent with the prevalence of edge states in the gap region of both \WS and \MS NPLs, the lowest-energy transitions primarily involve edge-localized states, with no core-originated transitions appearing below 4~eV in the spectrum of the smallest flake (0.9~nm). As the flake size increases to 2.1~nm, core-localized transitions progressively emerge but they remain systematically higher in energy compared to the edge-localized excitations, in agreement with the localization of the electronic states near the Fermi energy. While in both \MS and \WS 1.5~nm NPLs manifolds of weak core-originate transitions appear in Figure~\ref{fig5}~(a,b), in the nanostructures with 2.1~nm radius, the density of these excited states and their relative strength increases dramatically, contributing substantially to the optical activities of both NPLs above 2~eV.

{The observed size dependence of these transitions demands a thorough comparison with the excitonic properties of extended TMDs. In these materials, the exciton Bohr radius typically ranges from 1–2~nm, which is comparable to the lateral dimensions of our NPLs. While conventional effective-mass models might predict similar lateral confinement effects for both materials, our results highlight a qualitative divergence. In \MS, the transitions exhibit a ``particle-in-a-box'' behavior with a strong size sensitivity (130~meV blue shift), indicating that the excitons are indeed laterally confined within the core. In contrast, the optical response in \WS is dominated by edge-localized states, which are intrinsically less sensitive to the quantum dot size. Hence, their energies are effectively pinned by the edge potential, showing only a 20 meV shift and effectively suppressing the expected size dependence of the primary optical transition.}

These findings suggest that core-originated transitions will dominate over edge transitions in larger NSs, leading to longer-lived exciton decay~\cite{Frauendorf2024}, consistent with the measured decay times summarized in Table~\ref{tab1}. 
Direct comparison of \MS and \WS NPLs reveals markedly enhanced edge-state emission in \WS. In 2.1 nm \WS, the edge-originated excited states at the spectral onset (0.2--0.5~eV) are $\sim$10 orders of magnitude brighter than the higher-energy core transitions, while in \MS, the relative intensity of the two manifolds is comparable. This striking contrast signifies stronger optical coupling and radiative recombination in \WS, consistent with our measurements. {Physically, this is due to the larger spatial confinement and higher transition dipole moments of the $W$-dominated edge states. The larger spin-orbit coupling and more localized $5d$ orbitals of W, compared to Mo, enhance the separation between the edge-trap manifold and the core continuum. Consequently, \WS nanostructures exhibit higher intrinsic radiative rates at the edges (Figure~\ref{fig2}) where excitons possess relatively high oscillator strengths (Figure~\ref{fig5}b).} In contrast, in \MS, the comparable core- and edge-state intensities promote state hybridization and non-radiative quenching, which also agrees with the findings.

Furthermore, our work contextualizes different carrier trapping as compared to larger TMD nanosheets~\cite{Pippia2023}. In \MS\ and \WS\ NPLs, we identify that lateral edges control the optical response which host optically bright hole traps (majorly) from highly localized Mo/W $d$-orbitals.
{While the high oscillator strength of these states suggests the potential for sub-gap emission, consistent with sub-gap features often attributed to defect-bound excitons or trions in larger TMD nanosheets~\cite{Chow2015, Markus2025}, we do not observe PL in the smallest NPLs. This can be understood as a kinetic competition where, despite the bright nature of the edge transitions, the high density of surface states and vibrational modes in these high-surface-area structures provides dominant non-radiative decay channels. Consequently, the carrier population is quenched on a sub-picosecond timescale, effectively overshadowing the intrinsic optical activity of these edge states.}

{Finally, these DFT results provide a physical basis to rationalize the experimental decay components in Table~\ref{tab1}. The longest component, uniquely observed in the larger NSs, can be assigned to radiative recombination from core-originated states. In the smaller NPLs, where the optical response is dominated by edge-localized states, bright edge traps lead to the ultrafast quenching of the core-exciton population. Hence, the nanosecond radiative component vanishes in the NPL limit, where the maximized edge-to-core ratio ensures that the non-radiative trapping rate effectively outpaces radiative recombination. This correlation is further evidenced by the absence of detectable PL in NPLs, in contrast to the sizeable emission observed in the NSs.}

In summary, this study offers a comprehensive understanding of the exciton dynamics in colloidal \WS and \MS NPLs, revealing the atomistic origin of their quenched PL. We demonstrate edge-located hole traps to explain the ultrafast sub-picosecond exciton decay in laterally confined TMD NPLs, even without the introduction of sulfur vacancies. Our combined experimental and theoretical approach, identifies these traps as optically bright. With increasing NPL size, the core steadily dominates, reducing the contribution of edge states and resulting in longer-lived exciton decay, consistent with our experimental observations on NSs. Furthermore, a high density of localized states at the edges enables efficient charge transfer. This explains the exceptional photocatalytic activity of TMD nanostructures. 

Our work unravels not only the fundamental physics behind exciton quenching in colloidal TMDs but also opens a pathway for future design strategies, like post-synthesis chemical treatment or functionalization, to optimize the optical behavior of these systems. With the gained understanding of the importance of metal $d$-states and their influence of NPL size and edge shape, we can now explore new avenues to enhance PL quantum efficiency and catalytic performance. We anticipate that the trap states predicted here should also manifest in other TMD nanostructures. Future work based on {the solution of the Bethe-Salpeter equation, including all electron-electron and electron-hole correlation mechanisms properly describing excitons, will provide a quantitative characterization of the excited states and their dynamics. Likewise, the inclusion of explicit ligands will improve our understanding of the impact of functionalization on PL efficiency.}

\begin{acknowledgement}
CC acknowledges support from the DFG Collaborative Research Center 1375, project no. 398816777, subproject A08.
SV and CC acknowledge funding from the QuantERA II program under the European Union’s Horizon 2020 research and innovation program, within the EQUAISE project (Grant Agreement no. 101017733). SG acknowledges support from the Humboldt Internship Program. JL acknowledges funding of the TA measurements by the DFG under contract INST 37/1160-1 FUGG (project nr. 458406921), gratefully acknowledges additional funding by the DFG under the Excellence Strategy of the Cluster of Excellence PhoenixD (EXC 2122, Project ID 390833453) and by an Athene Grant of the University of Tübingen (by the Federal Ministry of Education and Research (BMBF) and the Baden-Württemberg Ministry of Science as part of the Excellence Strategy of the German Federal and State Governments). JCM and MK acknowledge funding from the German Research Foundation (DFG), project ME 3313/6-1, by the Ministry of Science, Research and Art Baden-Württemberg, Germany and by the European Union under EU-EFRE grant no. 712889.

\end{acknowledgement}

\begin{suppinfo}
The following files are available free of charge.
\begin{itemize}
 \item {si.pdf: The supporting information includes: Experimental and computational method details; structural properties, element and orbital resolved PDOS for all structures from main text.}
 
\end{itemize}

\end{suppinfo}
\vspace{0.5 cm}

\noindent\textbf{Data Availability Statement}\\
Input and output files of the ab initio calculations performed in this work are available free of charge in Zenodo \href{https://doi.org/10.5281/zenodo.17661456}{DOI: 10.5281/zenodo.17661456} [record: 17661456]. Experimental data is available from the corresponding author JL upon reasonable request.
\vspace{0.5 cm}

\noindent\textbf{Notes}\\
The authors declare no competing financial interest.

\bibliography{main}

\end{document}